\documentstyle[preprint,aps]{revtex}

\begin{document}

\draft

\tighten

\preprint{\vbox{\hfill chao-dyn/9603012 \\
          \vbox{\hfill UCSB--TH--96--05} \\ 
          \vbox{\hfill March 1996} \\
          \vbox{\vskip1.0in}
         }}

\title{Gaussian Fluctuations in Chaotic Eigenstates}

\author{Mark Srednicki\footnote{E--mail: \tt mark@tpau.physics.ucsb.edu}
   and Frank Stiernelof}

\address{Department of Physics, University of California,
         Santa Barbara, CA 93106 
         \\ \vskip0.5in}

\maketitle

\begin{abstract}
\normalsize{
We study the fluctuations that are predicted in the autocorrelation
function of an energy eigenstate of a chaotic, two-dimensional billiard
by the conjecture (due to Berry) that the eigenfunction is a gaussian 
random variable.  We find an explicit formula for the root-mean-square 
amplitude of the expected fluctuations in the autocorrelation function.
These fluctuations turn out to be $O(\hbar^{1/2})$
in the small $\hbar$ (high energy) limit.
For comparison, any corrections due to scars from isolated periodic orbits
would also be $O(\hbar^{1/2})$.  
The fluctuations take on a particularly simple form if
the autocorrelation function is
averaged over the direction of the separation vector.
We compare our various predictions with recent numerical computations of 
Li and Robnik for the Robnik billiard, and find good agreement.
We indicate how our results generalize to higher dimensions.}
\end{abstract}


Two-dimensional billiards which are classically chaotic have proven to
be an efficient laboratory for the study of quantum chaos.  
The energy eigenvalues and eigenfunctions can be computed
with good accuracy, and compared with theoretical
predictions of their properties.  These predictions are
typically semiclassical in nature, involving properties that
are expected to be emergent in the formal limit of $\hbar\to 0$.

In practice, numerical methods are used to find the
eigenvalues $k^2$ and eigenfunctions $\psi_k({\bf x})$
of the time-independent Schr\"odinger equation,
\begin{equation}
(\nabla^2 + k^2)\psi_k({\bf x}) = 0 \;, 
\label{schr}
\end{equation}
where $\bf x$ is in the domain $B$ of the billiard,
and the Dirichlet boundary condition
\begin{equation}
\psi_k({\bf x}) = 0 \quad \hbox{on} \quad \partial B 
\label{bc}
\end{equation}
is imposed.  Then large $k$ corresponds to small $\hbar$;
an expansion of some quantity in powers of the wavelength $\lambda = 2\pi/k$
corresponds to an expansion in powers of $\hbar$.

Our focus here will be on the autocorrelation function $C_{k,R}({\bf s})$,
introduced by Berry \cite{berry77b}.
Given an eigenfunction $\psi_k({\bf x})$,
a separation vector $\bf s$, and an averaging region $R$,
the autocorrelation function is defined to be
\begin{equation}
C_{k,R}({\bf s}) \equiv {A_B\over A_R}\int_R d^2x \;
\psi_k({\bf x}+{\textstyle{1\over2}}{\bf s}) 
\psi_k({\bf x}-{\textstyle{1\over2}}{\bf s}) \;,
\label{crs}
\end{equation}
where $A_B$ is the area of the billiard and $A_R$ is the area of the averaging
region $R$.  The eigenfunction $\psi_k({\bf x})$ is assumed to be real, and 
normalized in the usual way,
\begin{equation}
\int_B d^2x\;\psi_k^2({\bf x}) = 1 \;. 
\label{norm}
\end{equation}
However, our normalization of $C_{k,R}({\bf s})$ differs slightly from 
Berry's; we will discuss the reason for this later.

Berry conjectured that an energy eigenfunction
in a chaotic billiard would appear locally to be a superposition of
plane waves with random directions of the momenta and random phases. 
This implies that the expected value of $C_{k,R}({\bf s})$ is
\begin{eqnarray}
\langle C_{k,R}({\bf s})\rangle 
&=& {\int d^2p\;\delta({\bf p}^2-k^2)e^{i{\bf p}\cdot{\bf s}} \over
     \int d^2p\;\delta({\bf p}^2-k^2)} 
\nonumber \\
\noalign{\medskip}
&=& J_0(k s) \;,
\label{crsexp}
\end{eqnarray}
where $s=|{\bf s}|$, and $J_0(x)$ is a Bessel function.  We now know 
that there are corrections to this result (``scars'' \cite{heller})
associated with isolated periodic orbits in the classical billiard.
Assuming that the averaging region $R$ encompasses many wavelengths
in the direction perpendicular to each orbit giving a scar, then
the scar corrections to (\ref{crsexp}) are suppressed by $O(\hbar^{1/2})$.

In the limit that the number of superposed plane waves becomes infinite,
the central limit theorem tells us that the function $\psi_k({\bf x})$ 
can be treated as a gaussian random variable \cite{berry77b,ogh,mackauf88}.
This means that we have prior information
(in the sense used in Bayesian statistical analysis)
about $\psi_k({\bf x})$ which can be represented by a functional
probability distribution of the form
\begin{equation}
P(\psi_k) = {\cal N}\exp\left[-{1\over2}\int_B d^2x_1\int_B d^2x_2\ 
\psi_k({\bf x}_1)K({\bf x}_1,{\bf x}_2)\psi_k({\bf x}_2)
\right]\;.
\label{prob}
\end{equation}
Here $\cal N$ is a normalization constant, and
$K({\bf x}_1,{\bf x}_2)$ is the inverse of 
$A_B^{-1}J_0(k|{\bf x}_1-{\bf x}_2|)$ in the sense that
\begin{equation}
\int_B d^2x_2\,K({\bf x}_1,{\bf x}_2) J_0(k|{\bf x}_2-{\bf x}_3|)
= A_B \delta^2({\bf x}_1-{\bf x}_3)
\label{inv}
\end{equation}
with ${\bf x}_1$ and ${\bf x}_3$ restricted to $B$.
The angle brackets in (\ref{crsexp}) are then defined to represent
an average over the probability distribution (\ref{prob}).  Thus we have
\begin{equation}
\langle\psi_k({\bf x}_1)\psi_k({\bf x}_2)\rangle 
= A_B^{-1}J_0(k|{\bf x}_1-{\bf x}_2|) \;.
\label{2psi}
\end{equation}
Combining this with the definition (\ref{crs}) of $C_{k,R}({\bf s})$
gives $\langle C_{k,R}({\bf s})\rangle = J_0(ks)$,
in agreement with (\ref{crsexp}).
However, the probability distribution (\ref{prob}) also contains
information about the fluctuations of $C_{k,R}({\bf s})$ about 
$\langle C_{k,R}({\bf s})\rangle$.  Our goal is to study the
properties of these fluctuations.

Before proceeding, let us recall
that there is striking numerical evidence in favor of another
consequence of (\ref{prob}): specifically, the probability $P(\chi)d\chi$
that $\psi_k({\bf x})$ has a value between $\chi$ and $\chi+d\chi$
at a particular point $\bf x$ is given by
\begin{equation}
P(\chi) = (A_B/2\pi)^{1/2}\exp(-{\textstyle{1\over2}}A_B\chi^2) \;.
\label{prob2}
\end{equation}
This prediction can be tested by dividing the billiard into 
small pixels, and making a histogram of the value of the eigenfunction at
each pixel.  This was first done by MacDonald and Kaufman \cite{mackauf88}
in their study of eigenfunctions of the stadium billiard with
$k^2\!A_B \simeq 1.3 \times 10^4$.
More recently, Li and Robnik \cite{lirob} studied eigenfunctions
of the Robnik billiard \cite{robnik} with 
$k^2\!A_B \simeq 2.5 \times 10^6$,
and found excellent agreement with (\ref{prob2}).
Generally, the prediction (\ref{prob}) is expected to be valid
provided that distortions of the billiard boundary on the scale
of the wavelength $\lambda=2\pi/k$ do not permit the formation
of an integrable billiard \cite{mackauf88}.  

Our main tool in studying the fluctuations of
$C_{k,R}({\bf s})$ about $\langle C_{k,R}({\bf s})\rangle$ will be
the relation \cite{ogh,sred,eck}
\begin{eqnarray}
\langle
\psi_k({\bf x}_1)
\psi_k({\bf x}_2)
\psi_k({\bf x}_3)
\psi_k({\bf x}_4)
\rangle 
&=&
\langle
\psi_k({\bf x}_1)
\psi_k({\bf x}_2)
\rangle \langle
\psi_k({\bf x}_3)
\psi_k({\bf x}_4)
\rangle \nonumber \\
&& + \, \langle
\psi_k({\bf x}_1)
\psi_k({\bf x}_3)
\rangle \langle
\psi_k({\bf x}_2)
\psi_k({\bf x}_4)
\rangle \nonumber \\
&& + \, \langle
\psi_k({\bf x}_1)
\psi_k({\bf x}_4)
\rangle \langle
\psi_k({\bf x}_2)
\psi_k({\bf x}_3)
\rangle \;.
\label{4psi}
\end{eqnarray}
In particular, we consider the quantity
\begin{eqnarray}
\Delta_{k,R}({\bf s}_1,{\bf s}_2) &\equiv&
\Bigl\langle\Bigl[
C_{k,R}({\bf s}_1)-\langle C_{k,R}({\bf s_1})\rangle
\Bigr]\Bigl[
C_{k,R}({\bf s}_2)-\langle C_{k,R}({\bf s_2})\rangle
\Bigr]\Bigr\rangle  \nonumber \\
\noalign{\medskip}
&=&
\left\langle
C_{k,R}({\bf s}_1) C_{k,R}({\bf s}_2)
\right\rangle - 
\langle C_{k,R}({\bf s_1})\rangle
\langle C_{k,R}({\bf s_2})\rangle \;.
\label{deltadef}
\end{eqnarray}
$\Delta_{k,R}({\bf s},{\bf s})^{1/2}$ represents the root-mean-square 
discrepancy to be expected between
$C_{k,R}({\bf s})$ and $\langle C_{k,R}({\bf s})\rangle$ \cite{sred},
while $\Delta_{k,R}({\bf s}_1,{\bf s}_2)$ tells us whether the discrepancies
for ${\bf s}={\bf s}_1$ are correlated with those 
for ${\bf s}={\bf s}_2$, and whether this correlation is positive
or negative.

Let us note that quantities such as
$\langle|\psi_k({\bf x}_1)|^{2n}|\psi_k({\bf x}_2)|^{2m}\rangle$
have been computed previously, but with the angle brackets
representing an average over a random potential \cite{dot}.
This random-potential average was subsequently shown to be
equivalent to the average over the eigenfunction probability 
distribution $P(\psi_k)$ \cite{sred2}.

Returning to (\ref{deltadef}), we use the definition (\ref{crs}) of 
$C_{k,R}({\bf s})$ and the combinatoric property (\ref{4psi}) to get
\begin{eqnarray}
\Delta_{k,R}({\bf s}_1,{\bf s}_2) &=&
{A_B^2\over A_R^2}\int_R d^2x_1\int_R d^2x_2 \; 
\Bigl[ 
\langle\psi_k({\bf x}_1+{\textstyle{1\over2}}{\bf s}_1)
       \psi_k({\bf x}_2+{\textstyle{1\over2}}{\bf s}_2)\rangle
\langle\psi_k({\bf x}_1-{\textstyle{1\over2}}{\bf s}_1)
       \psi_k({\bf x}_2-{\textstyle{1\over2}}{\bf s}_2)\rangle
\nonumber \\
&& \qquad\qquad\qquad\qquad + \,
\langle\psi_k({\bf x}_1+{\textstyle{1\over2}}{\bf s}_1)
       \psi_k({\bf x}_2-{\textstyle{1\over2}}{\bf s}_2)\rangle
\langle\psi_k({\bf x}_1-{\textstyle{1\over2}}{\bf s}_1)
       \psi_k({\bf x}_2+{\textstyle{1\over2}}{\bf s}_2)\rangle 
\Bigr].
\nonumber \\
\label{delta1}
\end{eqnarray}
Now using (\ref{2psi}), we find 
\begin{eqnarray}
\Delta_{k,R}({\bf s}_1,{\bf s}_2) &=&
{1\over A_R^2}\int_R d^2x_1\int_R d^2x_2\;
\Bigl[
\, J_0(k|{\bf u} + {\bf s}_-|) \, J_0(k|{\bf u} - {\bf s}_-|)
\nonumber \\
&& \qquad\qquad\qquad\qquad +
\, J_0(k|{\bf u} + {\bf s}_+|) \, J_0(k|{\bf u} - {\bf s}_+|) \,
\Bigr] \;,
\label{delta2}
\end{eqnarray}
where we have defined
\begin{equation}
{\bf u} = {\bf x}_1 - {\bf x}_2  \qquad \hbox{and} \qquad
{\bf s}_\pm = {\textstyle{1\over2}}({\bf s}_1\pm{\bf s}_2) \;.
\label{us}
\end{equation}

To proceed further, we assume that the area $A_R$ is large, in the sense
that both
\begin{equation}
A_R \gg \lambda^2 \qquad \hbox{and} \qquad A_R \gg s^2_{1,2} \;.
\label{biga}
\end{equation}
In this case, the argument of each Bessel function is large over most
of the range of the integrand, and we can use the asymptotic formula
\begin{equation}
J_0(x) \simeq \left({2\over\pi x}\right)^{1/2}
\cos\Bigl(x-{\textstyle{\pi\over4}}\Bigr) \;,
\label{bigbess}
\end{equation}
which in fact is an excellent approximation for all $x > 1$.
Making the replacement (\ref{bigbess}), expanding in $s/u$,
and keeping only those terms which are not suppressed by extra powers
of either $ku$ or $s^2/u^2$, we have
\begin{eqnarray}
J_0(k|{\bf u} + {\bf s}|) J_0(k|{\bf u} - {\bf s}|) 
&\simeq& 
{2\over\pi ku}
\cos\Bigl(ku+ks\cos\theta-{\textstyle{\pi\over4}}\Bigr)
\cos\Bigl(ku-ks\cos\theta-{\textstyle{\pi\over4}}\Bigr)
\nonumber \\
\noalign{\medskip}
&\simeq& 
{1\over\pi ku}\Bigl[\sin(2ku) + \cos(2ks\cos\theta)\Bigr] \;,
\label{bessprod} 
\end{eqnarray}
where $\theta$ is the angle between $\bf u$ and $\bf s$.
We now use (\ref{bessprod}) in (\ref{delta2}), and notice that
the rapid oscillations of $\sin(2ku)$ will cause this term to
integrate to zero (to a good approximation).  Thus we find
\begin{eqnarray}
\Delta_{k,R}({\bf s}_1,{\bf s}_2) &=&
{1\over\pi k A_R^2}\int_R d^2x_1\int_R d^2x_2\; u^{-1}
\Bigl[\cos(2ks_-\cos\theta_-) + \cos(2ks_+\cos\theta_+) \Bigr] \;,
\label{delta3}
\end{eqnarray}
where $\theta_+$ ($\theta_-$) is the angle between $\bf u$ and 
${\bf s}_+$ (${\bf s}_-$).

To get a more explicit formula,
we need to choose the shape of the averaging region $R$.  For a disk of 
diameter $d$ and area $A_R=\frac14 \pi d^2$, the integrals in (\ref{delta3})
can be done in closed form by changing the integration variables to 
${\bf u} = {\bf x}_1 - {\bf x}_2$ and ${\bf v} = {\bf x}_1 + {\bf x}_2$,
integrating over $\bf v$ subject to the constraints
$|{\bf v} \pm {\bf u}| < d$, and then integrating over the magnitude
of $\bf u$ to get
\begin{equation}
\Delta_{k,R}({\bf s}_1,{\bf s}_2) 
= {16\over3\pi^{3/2}k A_R^{1/2}}\int_0^{2\pi}{d\theta\over 2\pi}\;
  \Bigl[\cos(2ks_-\cos\theta) + \cos(2ks_+\cos\theta) \Bigr]\;,
\label{delta4}
\end{equation}
where $\theta_+$ and $\theta_-$ have each been shifted and renamed $\theta$.
Performing the integral over $\theta$ gives us our central result,
\begin{equation}
\Delta_{k,R}({\bf s}_1,{\bf s}_2) 
= {16\over3\pi^{3/2}k A_R^{1/2}}
  \Bigl[J_0(k|{\bf s}_1-{\bf s}_2|) + J_0(k|{\bf s}_1+{\bf s}_2|) \Bigr] \;.
\label{delta5}
\end{equation}

We now turn to a study of the implications of (\ref{delta5}).  
The expected discrepancy between
$C_{k,R}({\bf s})$ and $\langle C_{k,R}({\bf s})\rangle$ 
is given by
\begin{equation}
\Delta_{k,R}({\bf s},{\bf s})^{1/2} = 1.38\,(k^2\!A_R)^{-1/4}
\left[{\textstyle{1\over2}}+{\textstyle{1\over2}}J_0(2ks)\right]^{1/2} \;,
\label{deltahalf}
\end{equation}
where the function in square brackets attains its maximum value of one 
when $ks=0$.  Since (\ref{deltahalf}) is proportional to $k^{-1/2}$, it is 
$O(\hbar^{1/2})$; thus, the RMS amplitude of the expected fluctuations
in $C_{k,R}({\bf s})$ vanishes in the classical limit.
However, this amplitude is not numerically small unless $A_R \gg \lambda^2$.
Both of these points are in accord with Berry's original (qualitative)
discussion of the approach of $C_{k,R}({\bf s})$ to 
$\langle C_{k,R}({\bf s})\rangle$ as $\hbar\to0$ \cite{berry77b}.
Furthermore, $\Delta_{k,R}({\bf s},{\bf s})^{1/2}$ is the same order
in $\hbar$ as any corrections due to scars.
This is consistent with the idea \cite{ogh} that scars represent a 
particular organization of the gaussian fluctuations in the eigenfunction,
rather than constituting an additional phenomenon.

For comparison, we turn to the numerical results of Li and Robnik
\cite{lirob} for the Robnik billiard \cite{robnik}.
We computed $C_{k,R}({\bf s})$ using the eigenfunction
with $k=790.644$, shown in fig.~(3) of \cite{lirob},
which was kindly supplied to us by Li and Robnik.
The averaging region $R$ was taken to be a disk with
diameter $d=0.273=34.4\,\lambda$.  
For this value of $d$, the coefficient of the bracketed function in 
(\ref{deltahalf}) is 0.100.  
The leading corrections to (\ref{delta5}) from terms we have neglected
(due to our various approximations) are suppressed by an extra factor of 
either $1/k A_R^{1/2} = 0.005$ or $s^2/A_R$; for $ks=30$, $s^2/A_R = 0.025$.
In figs.~(1--12), we plot the actual correlation
function $C_{k,R}({\bf s})$ as a solid line, along with
a shaded band encompassing the range
$\langle C_{k,R}({\bf s})\rangle \pm \Delta_{k,R}({\bf s},{\bf s})^{1/2}$.
The inset in each figure depicts the Robnik billiard, and the filled
circle shows the averaging region which was used.
The double-headed arrow in each inset has unit length, 
and its direction shows the direction of $\bf s$.
We see that the actual $C_{k,R}({\bf s})$ always lies within the 
shaded band for a majority of the time, but does have (sometimes large)
excursions outside of it.
Without attempting a detailed quantitative analysis, we can say
that these graphs are qualitatively consistent with what we expect.

Li and Robnik \cite{lirob} suggested that the discrepancy between
$C_{k,R}({\bf s})$ and $\langle C_{k,R}({\bf s})\rangle$
could be reduced by averaging $C_{k,R}({\bf s})$
over the direction of $\bf s$.  
Let us define
\begin{equation}
{\bar C}_{k,R}(s) \equiv \int_0^{2\pi} {d\phi\over2\pi}\;
C_{k,R}({\bf s}(\phi)) \;,
\label{cav}
\end{equation}
where ${\bf s}(\phi)=(s\cos\phi,\, s\sin\phi)$.  Obviously, we have
\begin{equation}
\langle {\bar C}_{k,R}(s)\rangle = \langle C_{k,R}({\bf s})\rangle 
    = J_0(ks) \;.
\label{cavexp}
\end{equation}
We also define
\begin{eqnarray}
{\bar\Delta}_{k,R}(s_1,s_2) &\equiv&
 \langle {\bar C}_{k,R}(s_1) {\bar C}_{k,R}(s_2) \rangle
-\langle {\bar C}_{k,R}(s_1) \rangle \langle {\bar C}_{k,R}(s_2) \rangle \;.
\nonumber \\
\noalign{\medskip}
&=& \int_0^{2\pi} {d\phi_1\over2\pi} \int_0^{2\pi} {d\phi_2\over2\pi}\;
\Delta_{k,R}({\bf s}_1(\phi_1), {\bf s}_2(\phi_2)) \;.
\label{deltaav1}
\end{eqnarray}
Then, using (\ref{delta5}), we find 
\begin{equation}
{\bar\Delta}_{k,R}(s_1,s_2) =
{32\over3\pi^{3/2}k A_R^{1/2}} \int_0^{2\pi} {d\phi\over2\pi}\;
     J_0(k[s_1^2 + s_2^2 \pm 2s_1s_2\cos\phi]^{1/2}) \;,
\label{deltaav2}
\end{equation}
where $\phi=\phi_1-\phi_2$; this integral can be performed to yield
\begin{equation}
{\bar\Delta}_{k,R}(s_1,s_2) 
           = {32\over3\pi^{3/2}k A_R^{1/2}}\, J_0(ks_1) J_0(ks_2) \;.
\label{deltaav3}
\end{equation}

The fact that ${\bar\Delta}_{k,R}(s_1,s_2)$ is proportional to
$\langle {\bar C}_{k,R}(s_1) \rangle \langle {\bar C}_{k,R}(s_2) \rangle$
has dramatic consequences; it implies that
${\bar C}_{k,R}(s) / \langle {\bar C}_{k,R}(s) \rangle$
must be independent of $s$.  To demonstrate this, we choose
a set of orthonormal basis functions $f_n(s)$, $n=0,1,2,\ldots\,$, 
with $f_0(s)$ chosen to be equal to $\langle{\bar C}_{k,R}(s)\rangle$.
We require orthonormality in the sense that
\begin{equation}
\int_0^\infty ds\,w(s)f_n(s)f_m(s) = \delta_{nm} \;,
\label{orthon}
\end{equation}
where $w(s)$ is any weight function which ensures the convergence
and correct normalization of the integral when $n=m=0$.  
(Since $\langle{\bar C}_{k,R}(s)\rangle = J_0(ks)$,
we could construct such a set of basis functions
by starting with the Bessel functions
$J_n(ks)$ and then performing Gram-Schmidt orthogonalization.)
Once we have the basis functions, we can write
\begin{eqnarray}
{\bar C}_{k,R}(s) &=& \sum_{n=0}^\infty c_n f_n(s) \;, \nonumber \\
\noalign{\medskip}
c_n &=& \int_0^\infty ds\,w(s)f_n(s) {\bar C}_{k,R}(s) \;,
\label{coeff}
\end{eqnarray}
where the $c_n$'s should be regarded as random variables.
By construction, we have
\begin{equation}
\langle c_n\rangle = \delta_{n0} \;. 
\label{cnexp}
\end{equation}
Using (\ref{deltaav3}), we can also compute the expected value of
$c_n^2$.  We find
\begin{eqnarray}
\langle c_n^2\rangle - \langle c_n\rangle^2 &=& 
\int_0^\infty ds_1\,w(s_1)f_n(s_1)
\int_0^\infty ds_2\,w(s_2)f_n(s_2)
{\bar\Delta}_{k,R}(s_1,s_2) \nonumber \\
\noalign{\medskip}
&=& {32\over3\pi^{3/2}k A_R^{1/2}}
\int_0^\infty ds_1\,w(s_1)f_n(s_1)f_0(s_1)
\int_0^\infty ds_2\,w(s_2)f_n(s_2)f_0(s_2) \nonumber \\
\noalign{\medskip}
&=& {32\over3\pi^{3/2}k A_R^{1/2}}\,\delta_{n0} \;.
\label{c2nexp}
\end{eqnarray}
Thus $\langle c_n^2\rangle=0$ if $n\ne0$, indicating that the
probability distribution has no support for any nonzero $c_n$ other
than $c_0$.  Therefore
${\bar C}_{k,R}(s) \propto f_0(s) = \langle {\bar C}_{k,R}(s) \rangle$.

However, we must remember that there are additional contributions
to ${\bar\Delta}_{k,R}(s_1,s_2)$ which are suppressed by an extra 
factor of either $1/k A_R^{1/2}$ or $s^2/A_R$,
and that these will make small corrections
to the functional form of ${\bar\Delta}_{k,R}(s_1,s_2)$.
This means that ${\bar C}_{k,R}(s) / \langle{\bar C}_{k,R}(s)\rangle$
should be independent of $s$ up to corrections of order
$1/k A_R^{1/2}$ and $s^2/A_R$.

The discrepancy between
${\bar C}_{k,R}(s)$ and $\langle{\bar C}_{k,R}(s)\rangle$
is governed by ${\bar\Delta}_{k,R}(s,s)^{1/2}$.  In figs.~(13--16),
we plot the actual direction-averaged correlation function 
${\bar C}_{k,R}(s)$ as a solid line, using the same four averaging
regions as before.  We also plot a shaded band encompassing the range
$\langle{\bar C}_{k,R}(s)\rangle \pm {\bar\Delta}_{k,R}(s,s)^{1/2}$.
We see that the actual ${\bar C}_{k,R}(s)$ is consistent with
our expectations.  In fig.~(17), we plot 
$[{\bar C}_{k,R}(s)-{\bar C}_{k,R}(0)J_0(ks)] + {\bar C}_{k,R}(0)$
for the four averaging regions; this quantity 
should be independent of $s$ and equal to ${\bar C}_{k,R}(0)$.
(We plot it instead of the ratio ${\bar C}_{k,R}(s)/J_0(ks)$ because 
the latter is dominated by numerical errors near the zeros of its
denominator.)
The plots are remarkably flat;
the small glitches which are present are most likely due to the build-up
of round-off errors in the numerical computation.
These plots confirm our
prediction that ${\bar C}_{k,R}(s) / \langle{\bar C}_{k,R}(s)\rangle$
should be independent of $s$.
This result is an incisive test of the validity of (\ref{deltaav3}),
and by implication, (\ref{4psi}).

This concludes our analysis of the fluctuations in the autocorrelation
function for the case of a circular averaging region.  We now consider
the dependence of $\Delta_{k,R}({\bf s}_1,{\bf s}_2)$
on the shape of the averaging region $R$. 

For noncircular $R$, the integrals in (\ref{delta3}) cannot be done
in closed form for nonzero ${\bf s}_{1,2}$.  For the special case 
${\bf s}_1={\bf s}_2=0$, however, we can evaluate (\ref{delta3})
for a rectangular averaging region.
This will give us the shape dependence of 
$\Delta_{k,R}({\bf 0},{\bf 0})$, and therefore (we hope)
some idea of the shape dependence of
$\Delta_{k,R}({\bf s}_1,{\bf s}_2)$ for general ${\bf s}_1$ and ${\bf s}_2$.
For a rectangle with edge lengths $a$ and $b$, each of which is much
greater than the wavelength $\lambda=2\pi/k$, we find
\begin{equation}
\Delta_{k,R}({\bf 0},{\bf 0}) 
= {4\over3\pi k A_R^{1/2}} 
  \Bigl[g(\xi)+g(\xi^{-1})-(\xi+\xi^{-1})^{3/2}\Bigr] \;,
\label{deltarect}
\end{equation}
where $A_R=ab$ is the rectangle's area, $\xi=b/a$ is the ratio of 
edge lengths, and
\begin{equation}
g(\xi) = \xi^{-3/2} + 3 \xi^{-1/2} \sinh^{-1}\xi \;.
\label{g}
\end{equation}
In fig.~(18), we plot the ratio of $\Delta_{k,R}({\bf 0},{\bf 0})$ for
a rectangle to $\Delta_{k,R}({\bf 0},{\bf 0})$ for a circle
of the same area, as a function of the edge length ratio of the 
rectangle.  We see that $\Delta_{k,R}({\bf 0},{\bf 0})$ exhibits
only a mild shape dependence.

Earlier numerical computations of 
$C_{k,R}({\bf s})$ \cite{mackauf88,lirob,aurstein93} 
are all in qualitative agreement with our considerations;
specifically, the average discrepancy between $C_{k,R}({\bf s})$ and
$\langle C_{k,R}({\bf s})\rangle$ is always roughly given by 
$(k^2\! A_R)^{-1/4}$.  A detailed comparison is hindered by two issues.
First, in accord with Berry's original definition, 
earlier authors usually work with an autocorrelation function
${\widetilde C}_{k,R}({\bf s})$ which, in our notation, is
\begin{equation}
{\widetilde C}_{k,R}({\bf s}) = { C_{k,R}({\bf s}) \over C_{k,R}({\bf 0})} \;.
\label{ctilde}
\end{equation}
From our point of view, ${\widetilde C}_{k,R}({\bf s})$ is a much more
complicated object than $C_{k,R}({\bf s})$; there is no simple expression
for $\langle{\widetilde C}_{k,R}({\bf s})\rangle$, because the
relevant functional integral (of $C_{k,R}({\bf s})/C_{k,R}({\bf 0})$ times
the probability distribution $P(\psi_k)$ over $\psi_k$) cannot be done
using the simple gaussian combinatorics of (\ref{4psi}).  Of course,
by definition ${\widetilde C}_{k,R}({\bf 0})=1$, and so essentially
what happens is that any fluctuation in $C_{k,R}({\bf 0})$ shows up as 
a multiplicative enhancement or suppression of fluctuations in 
${\widetilde C}_{k,R}({\bf s})$ at nonzero $\bf s$.  

Another problem occurs if an axis of symmetry of the billiard passes
through the averaging region.  Every energy eigenfunction is either
symmetric or antisymmetric under reflection about such an axis;
this can be handled analytically by writing
\begin{equation}
\psi_k(x,y)={1\over\sqrt2}\Bigl[\chi_k(x,y) + \chi_k(x,-y)\Bigr]
\label{sym}
\end{equation}
(where we have illustratively assumed an eigenfunction symmetric about the
$x$-axis), and treating $\chi_k({\bf x})$ as a gaussian random variable.  
This approach considerably enhances the complexity of the analysis, however;
for example, the number of independent terms on the right-hand-side of 
(\ref{4psi}) grows from three to forty-eight.  The simplest solution is 
to do numerical analysis with averaging regions that do not cross any axes
of symmetry, as we have done here.

Finally, we note that all of our results have a straightforward 
generalization to higher dimensions.  For a $D$-dimensional
billiard, the autocorrelation function becomes \cite{berry77b}
\begin{eqnarray}
\langle C_{k,R}({\bf s})\rangle 
&=& {\int d^Dp\;\delta({\bf p}^2-k^2)e^{i{\bf p}\cdot{\bf s}} \over
     \int d^Dp\;\delta({\bf p}^2-k^2)} 
\nonumber \\
\noalign{\medskip}
&=& 2^{(D-2)/2}\Gamma(D/2)\,{J_{(D-2)/2}(k s) \over (k s)^{(D-2)/2} }
\nonumber \\
\noalign{\medskip}
&\equiv& F_D(k s) \;,
\label{crsexpd}
\end{eqnarray}
where $J_\nu(x)$ is a Bessel function.  The generalizations of
(\ref{delta5}) and (\ref{deltaav3}), which follow from the properties
of $F_D(k s)$, are
\begin{equation}
\Delta_{k,R}({\bf s}_1,{\bf s}_2) 
= {\gamma_D\over k_{\vphantom{R}}^{\vphantom{(}D-1} V_R^{(D-1)/D} }
  \Bigl[F_D(k|{\bf s}_1-{\bf s}_2|) + F_D(k|{\bf s}_1+{\bf s}_2|) \Bigr] 
\label{delta5d}
\end{equation}
and
\begin{equation}
{\bar\Delta}_{k,R}(s_1,s_2) 
= {2\gamma_D\over k_{\vphantom{R}}^{\vphantom{(}D-1} V_R^{(D-1)/D} } 
  \, F_D(ks_1) F_D(ks_2) \;,
\label{deltaav3d}
\end{equation}
where $V_R$ is the $D$-dimensional volume of the spherical averaging 
region, and $\gamma_D$ is a numerical factor which we have not computed.

To conclude, we have performed an analysis of the autocorrelation function
$C_{k,R}({\bf s})$ under the assumption that the energy eigenfunction
$\psi_k({\bf x})$ behaves like a gaussian random variable,
in a sense which we have made precise.
We find that, for a two-dimensional billiard, 
$C_{k,R}({\bf s})$ should have $O(\hbar^{1/2})$ fluctuations
about its expected value $\langle C_{k,R}({\bf s})\rangle = J_0(ks)$;
scars from isolated periodic orbits would give corrections to 
$C_{k,R}({\bf s})$ which are also $O(\hbar^{1/2})$.
We have given analytic formulae for the root-mean-square
amplitude of the expected fluctuations in $C_{k,R}({\bf s})$.
We find that a particularly useful object to study is 
${\bar C}_{k,R}(s)$, which is $C_{k,R}({\bf s})$ averaged over the
angle of $\bf s$.  We predict that ${\bar C}_{k,R}(s)/J_0(ks)$ is
independent of $s$, a prediction which is very well satisfied by
the numerical results of Li and Robnik for the Robnik billiard.

\begin{acknowledgments}

We are grateful to Baowen Li and Marko Robnik for kindly providing
us with some of their data for the Robnik billiard. 
We thank Eric Heller and Steven Tomsovic for helpful discussions.
This work was supported in part by NSF Grant PHY--91--16964.

\end{acknowledgments}

\vskip0.5in

\centerline{\bf FIGURE CAPTIONS}

\vskip0.2in

Fig.~1~~The autocorrelation function $C_{k,R}({\bf s})$ is shown as a 
solid line. The gray band depicts 
$\langle C_{k,R}({\bf s})\rangle \pm \Delta_{k,R}({\bf s},{\bf s})^{1/2}$,
which is the expected root-mean-square range of $C_{k,R}({\bf s})$.
The Robnik billiard is shown in the inset;
the averaging region $R$ is indicated by the filled circle,
and the direction of the separation vector $\bf s$ is indicated 
by the direction of the two-headed arrow, which has unit length.  

\vskip0.2in

Figs.~2--12~~Same as fig.\ (1).

\vskip0.2in

Fig.~13~~The autocorrelation function ${\bar C}_{k,R}(s)$, averaged over
the direction of the separation vector $\bf s$, is shown as a solid line.
The gray band depicts $\langle {\bar C}_{k,R}(s)\rangle \pm 
{\bar\Delta}_{k,R}(s,s)^{1/2}$,
which is the expected root-mean-square range of ${\bar C}_{k,R}(s)$.
The Robnik billiard is shown in the inset;
the averaging region $R$ is indicated by the filled circle.

\vskip0.2in

Figs.~14--16~~Same as fig.\ (13).

\vskip0.2in

Fig.~17~~$[{\bar C}_{k,R}(s) - {\bar C}_{k,R}(0)J_0(ks)] + {\bar C}_{k,R}(0)$
for each of the four averaging regions.
This quantity is predicted to be independent of $s$. 

\vskip0.2in

Fig.~18~~The ratio of $\Delta_{k,R}({\bf 0},{\bf 0})$ for a rectangle
with edge length ratio $b/a$ to
$\Delta_{k,R}({\bf 0},{\bf 0})$ for a circle of the same area.

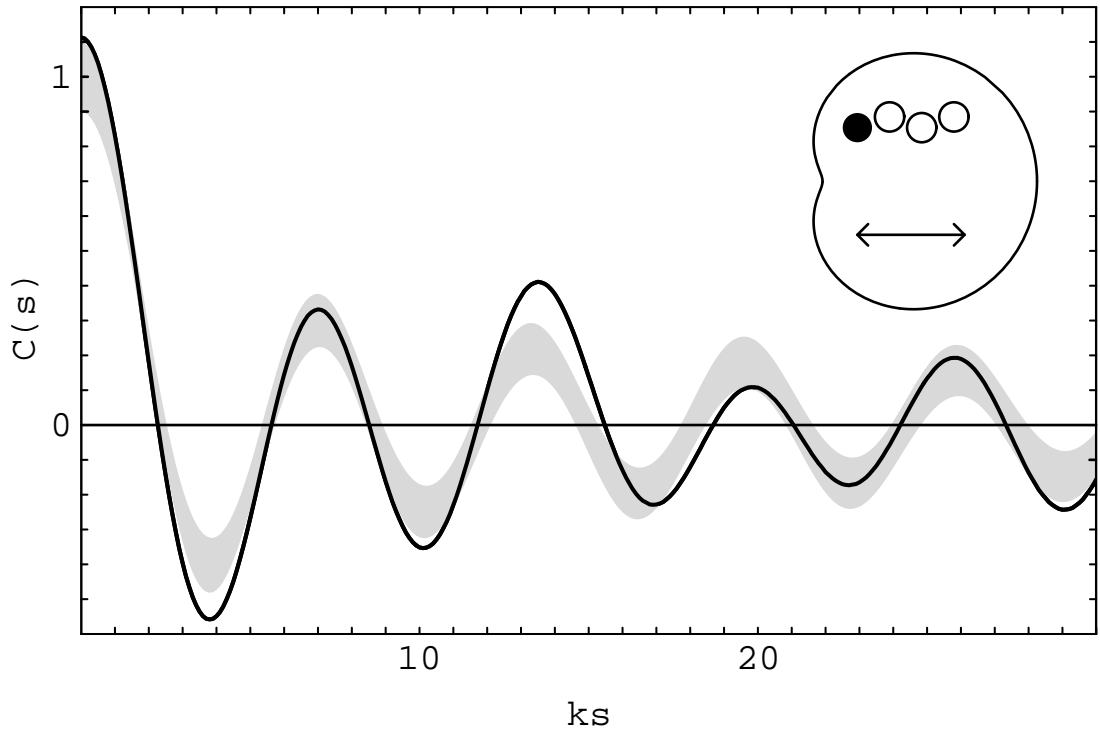
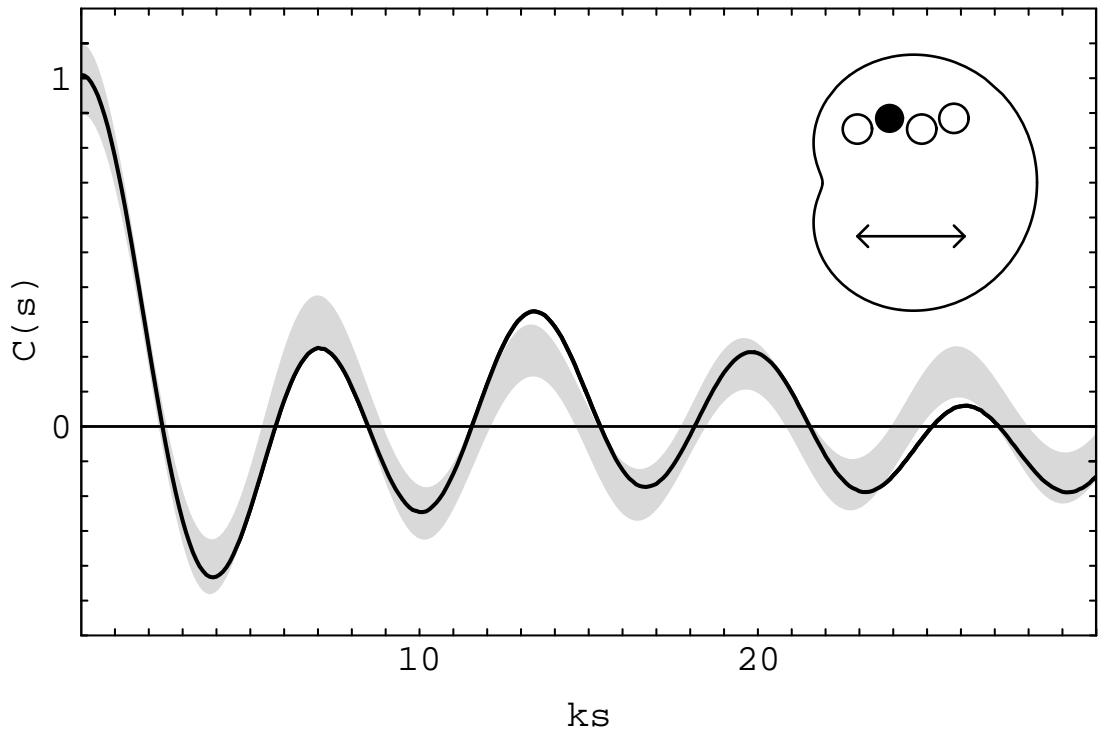
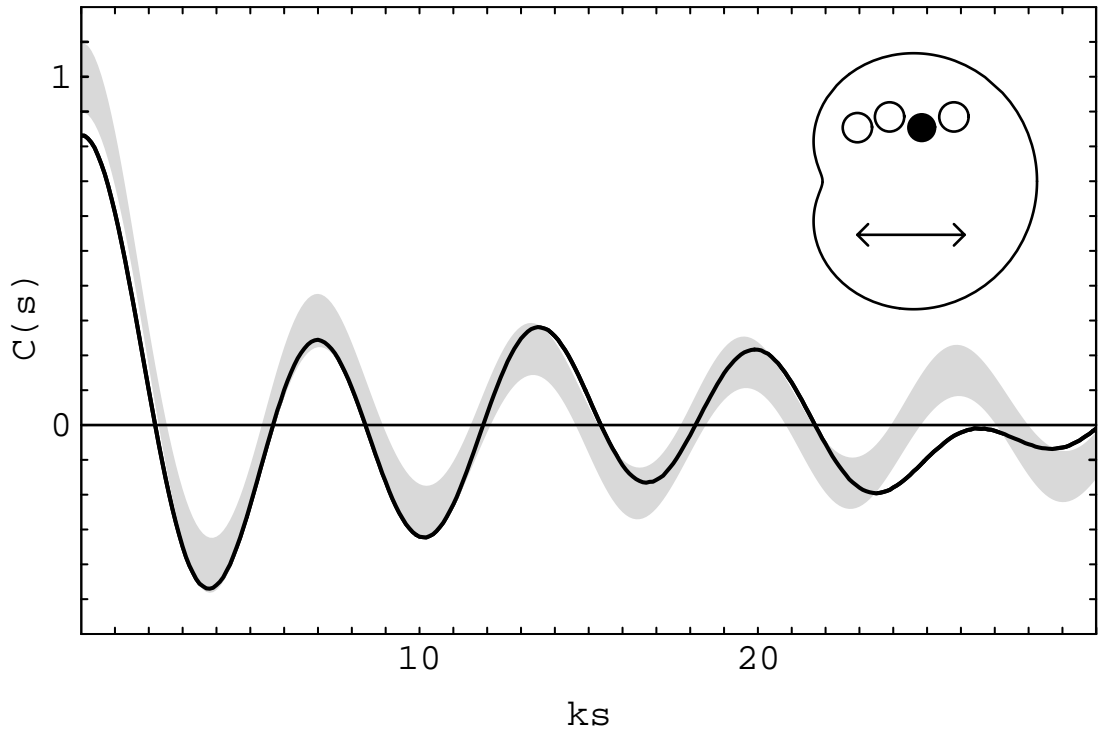
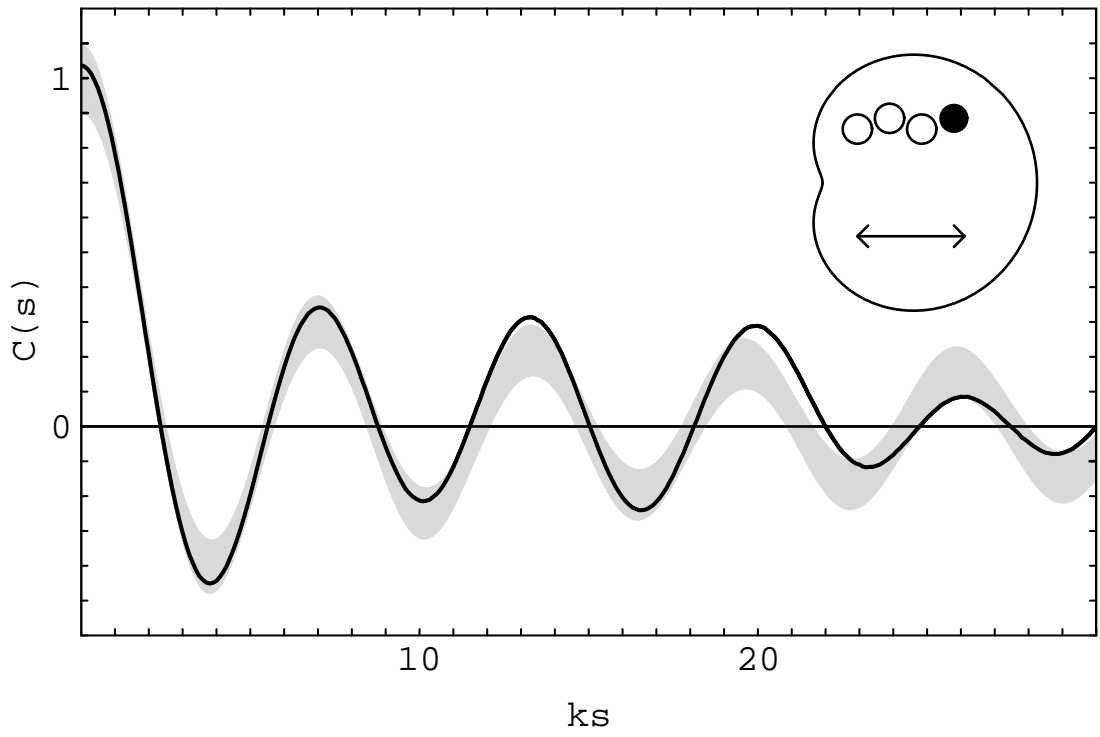
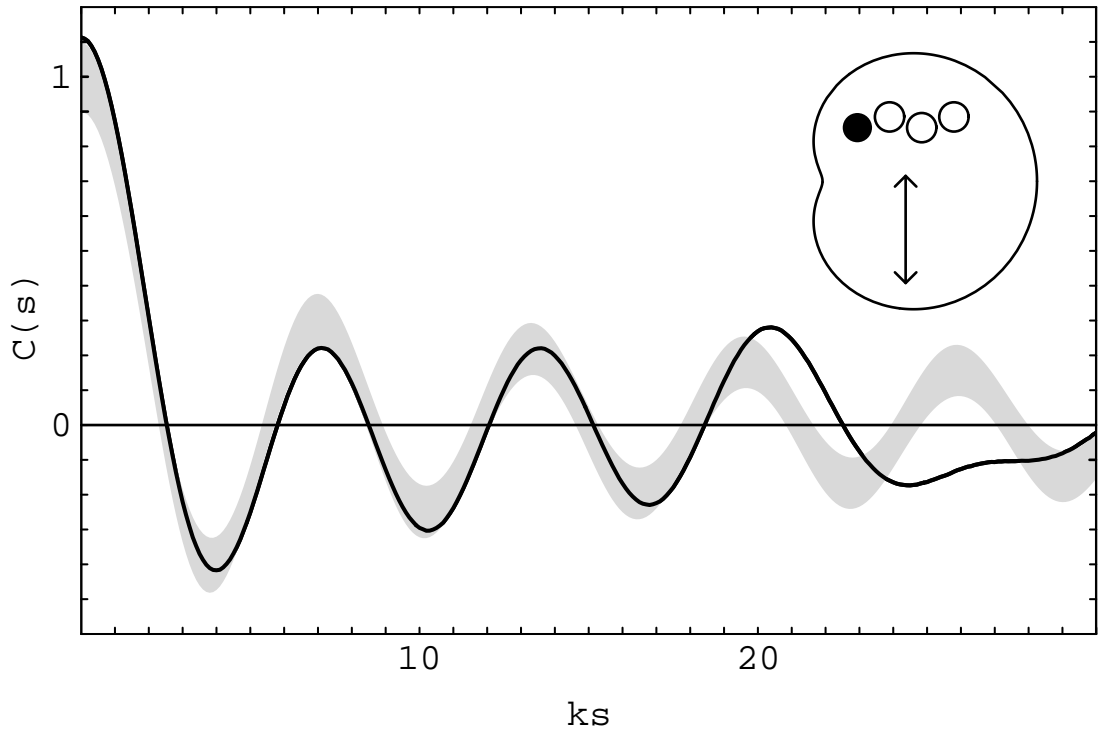
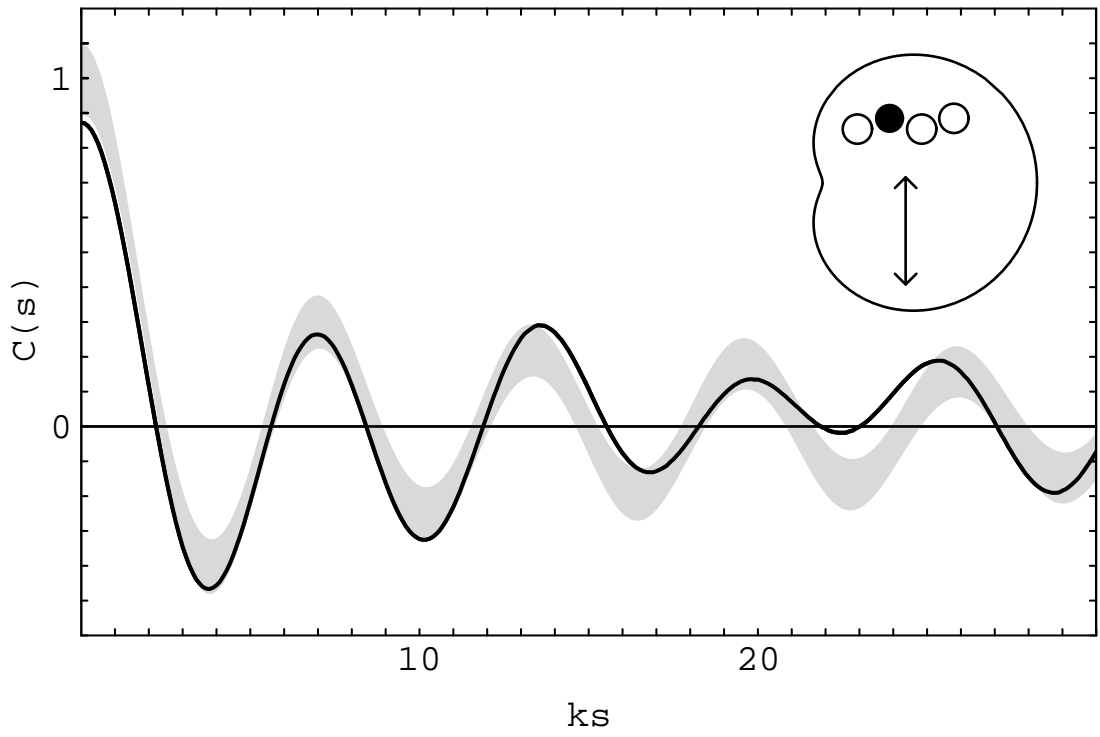
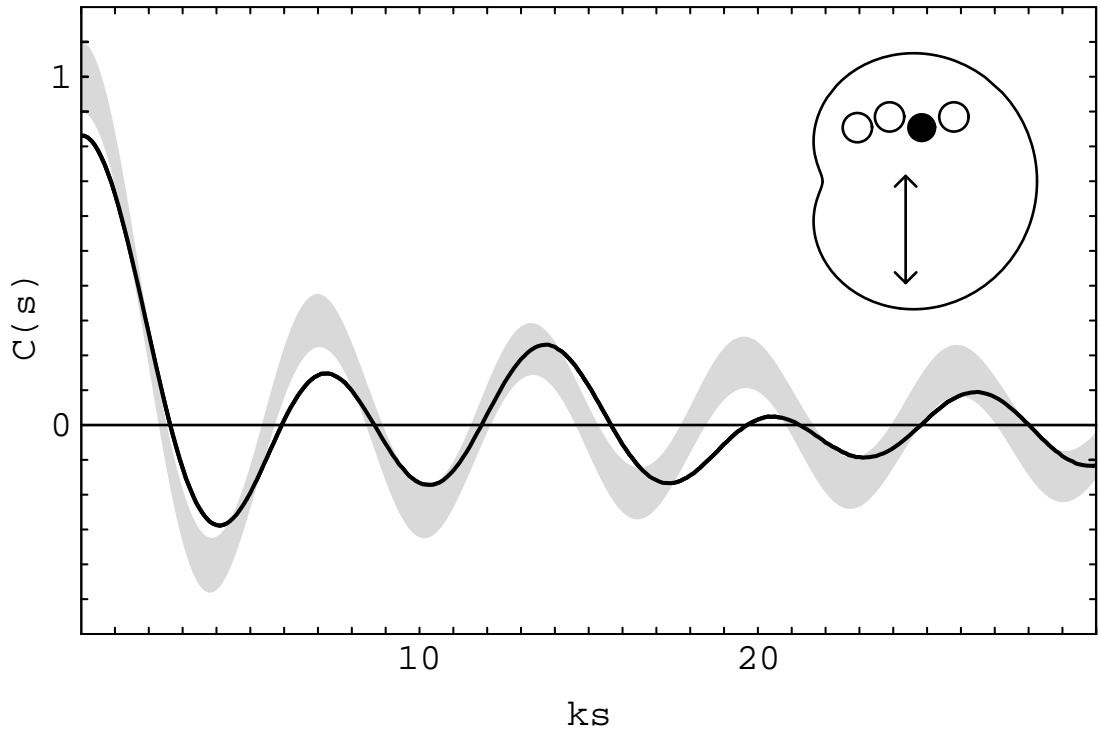
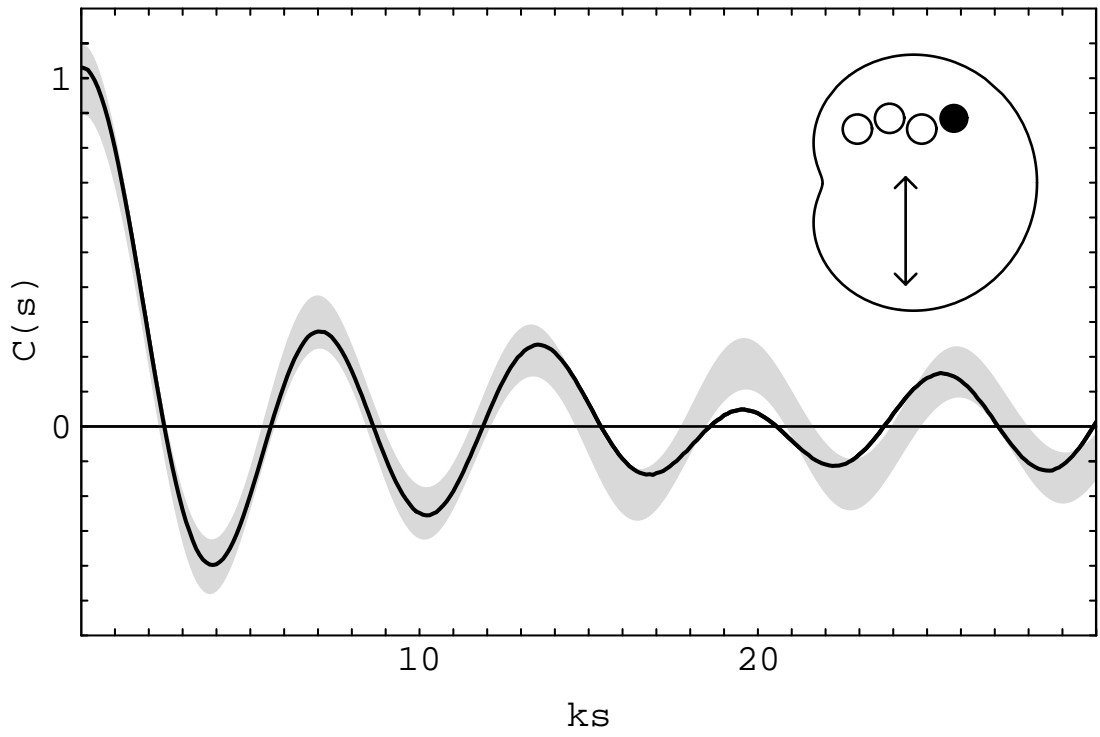
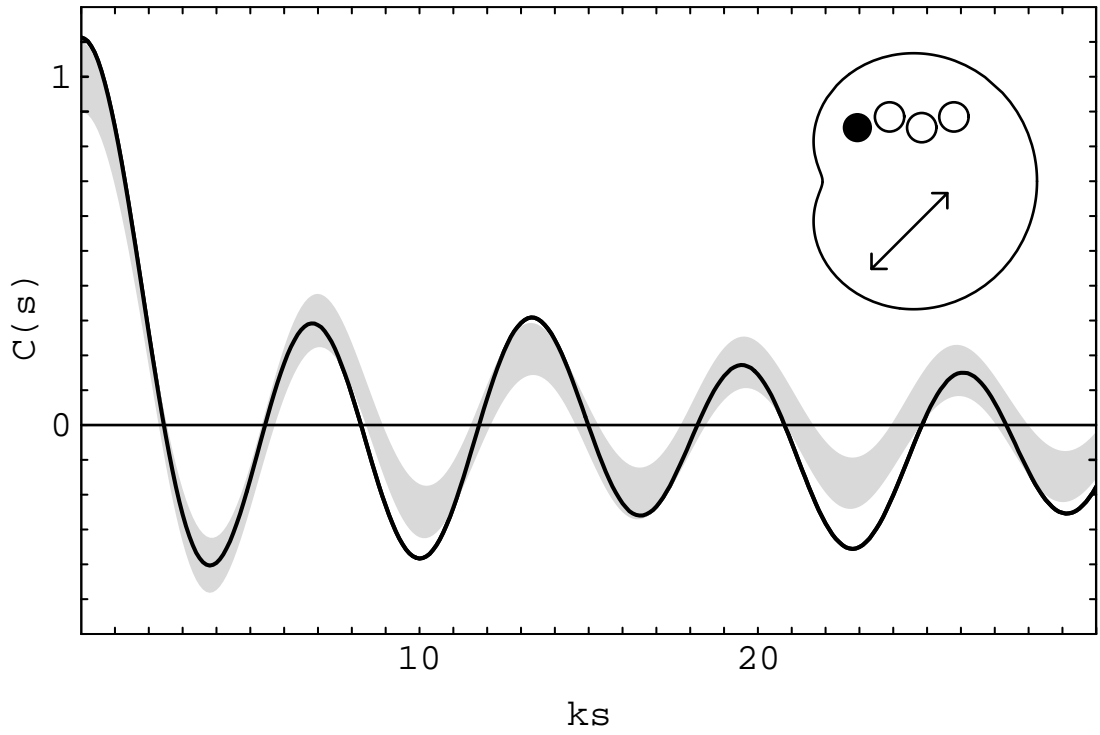
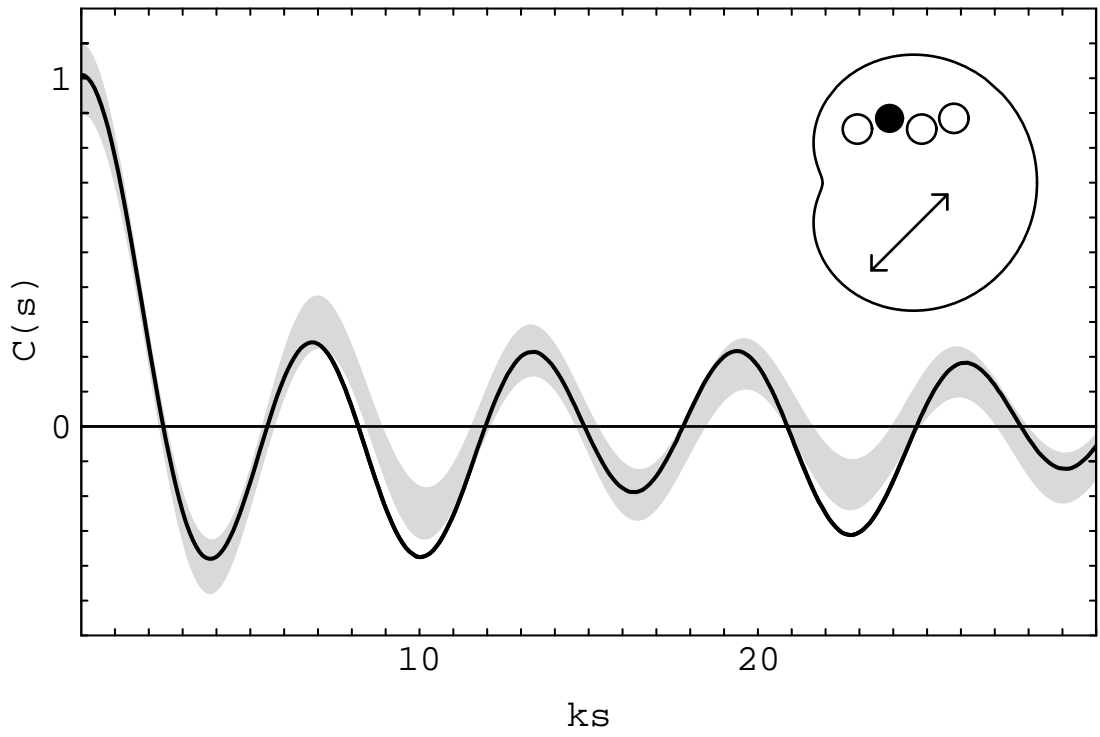
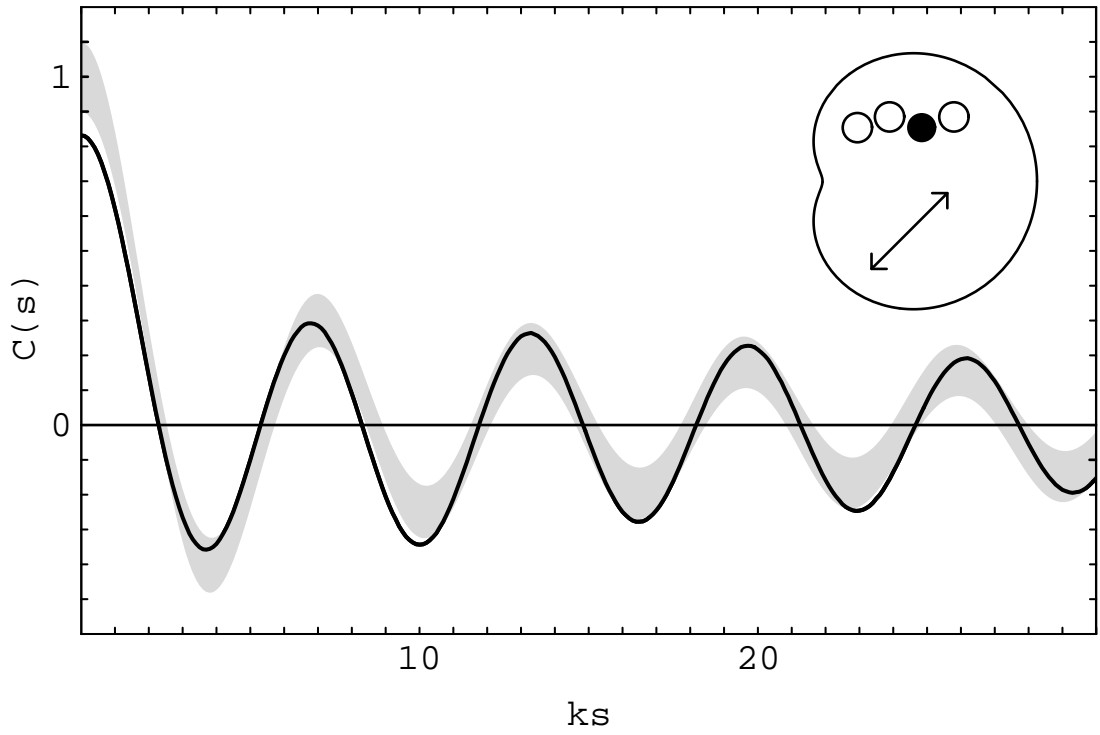
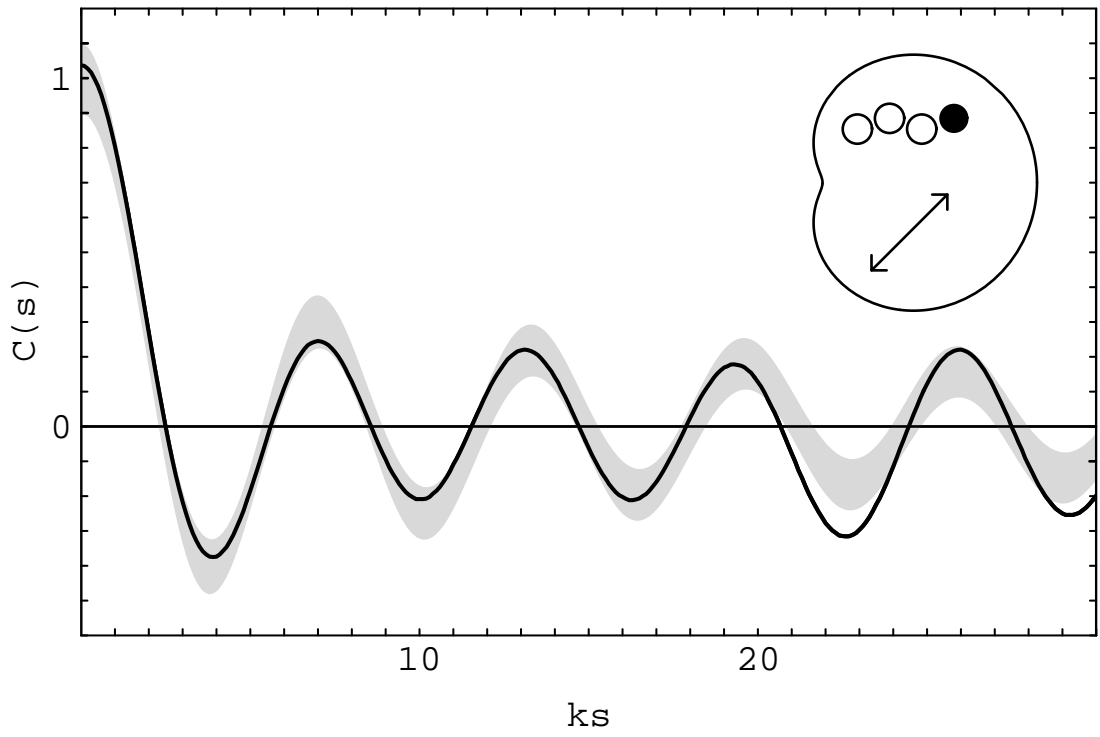
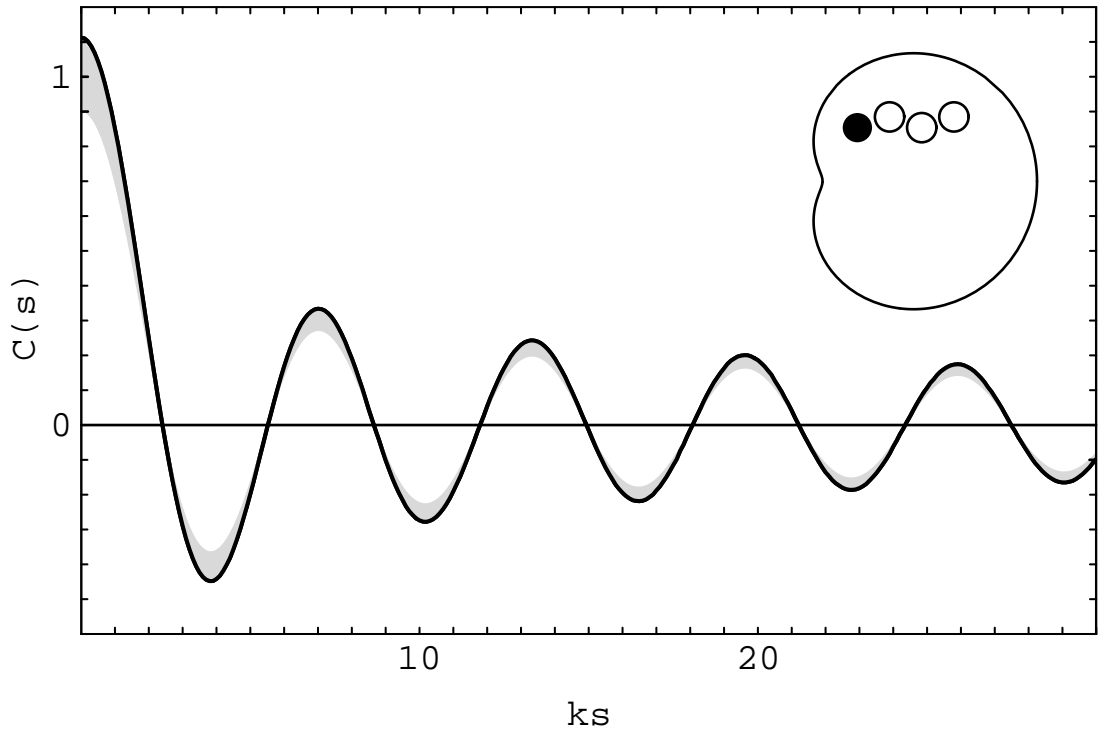
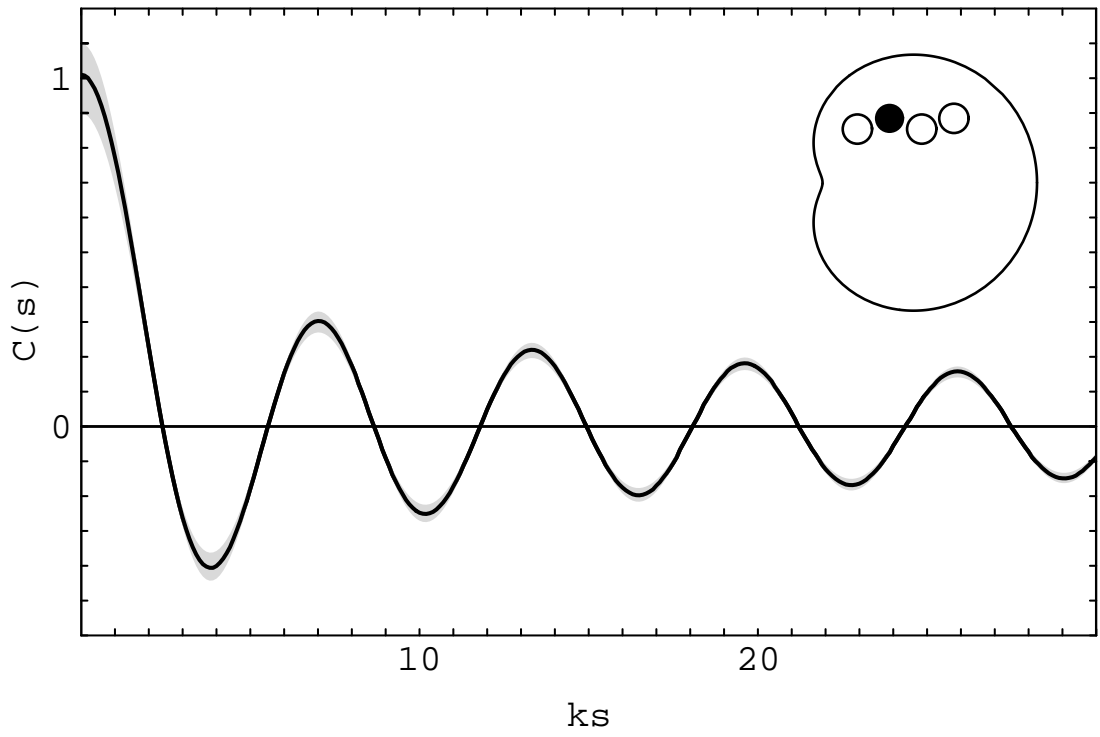
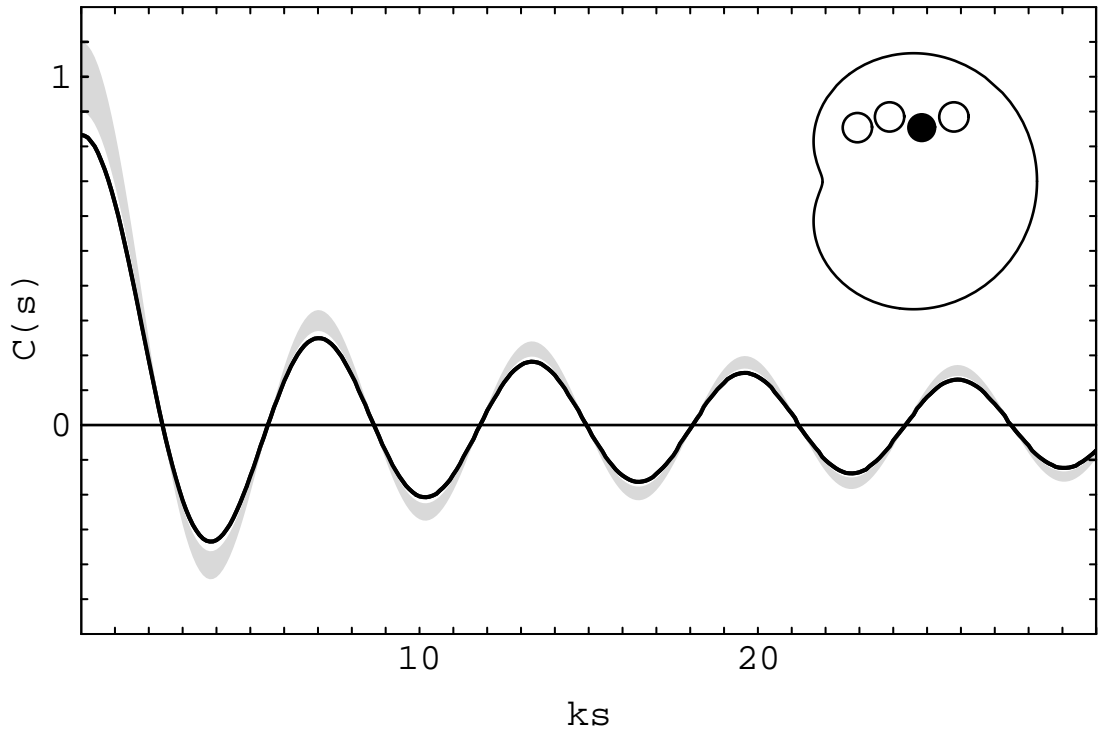
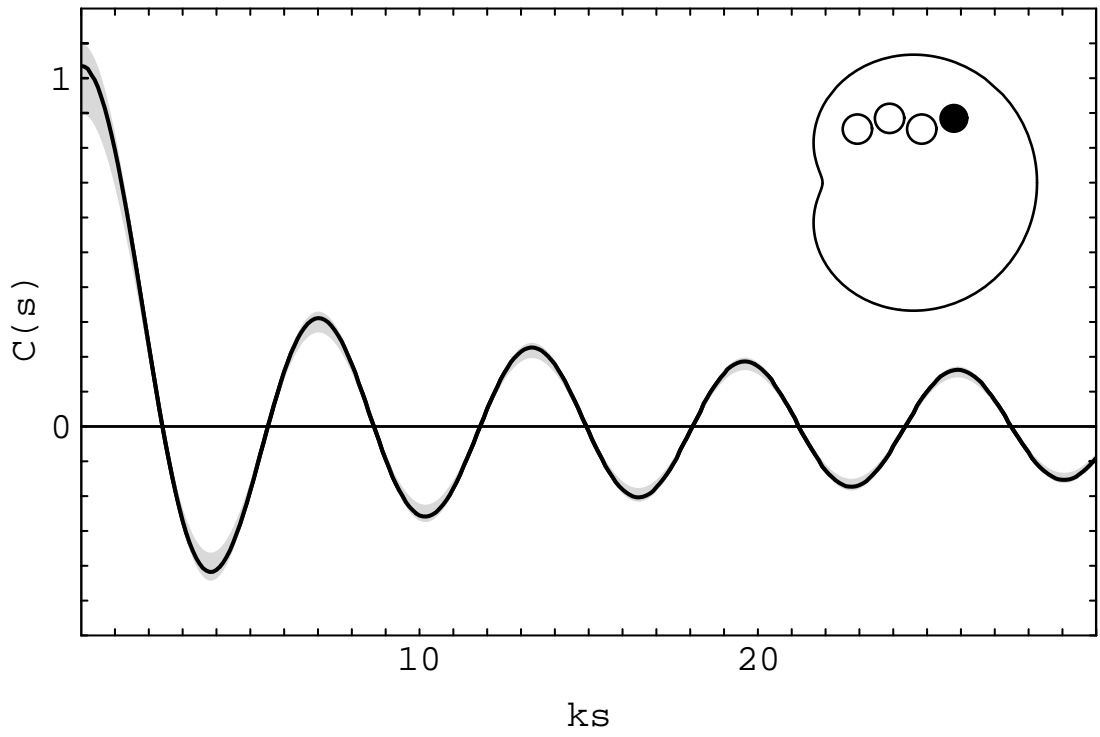
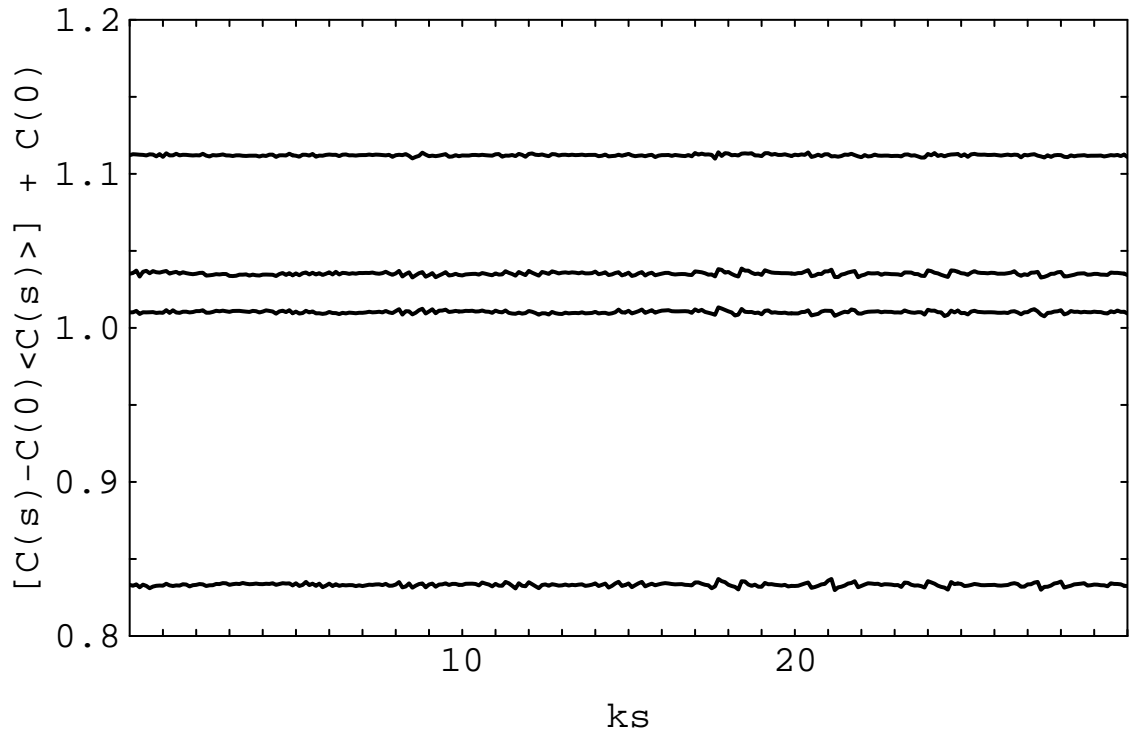
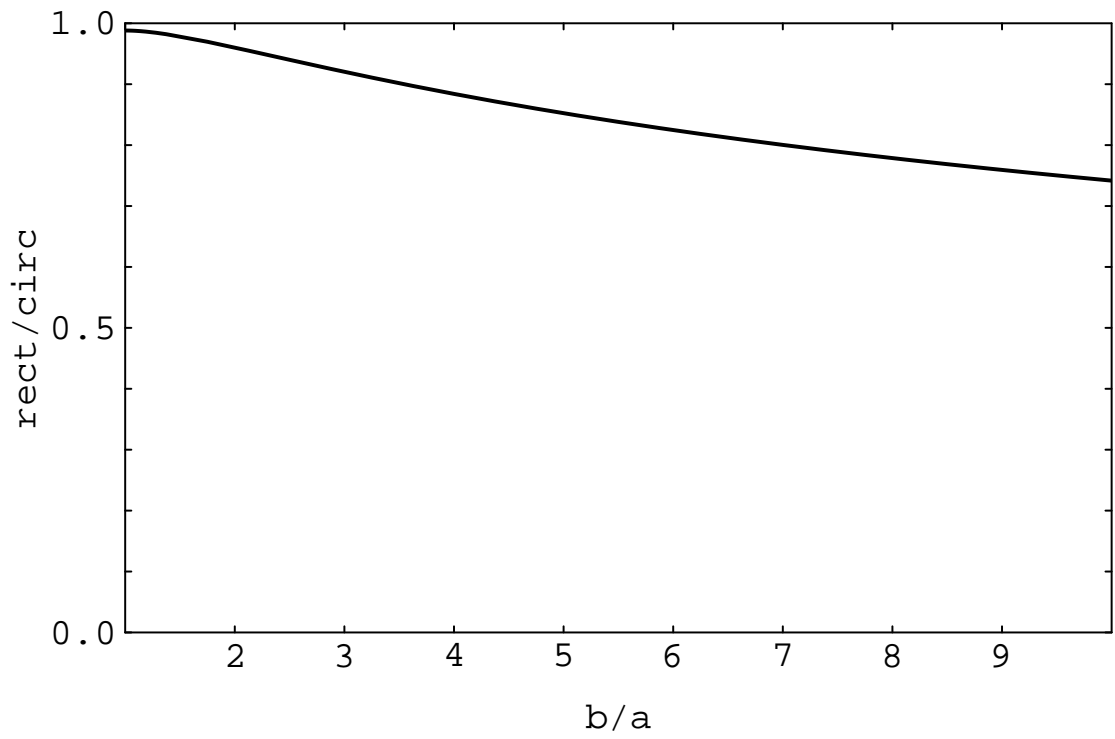
\end{document}